\documentclass{appolb}
\usepackage{amsfonts}
\usepackage{amssymb}
\usepackage{amsmath}
\usepackage{soul}
\usepackage{epsfig}
\usepackage{graphicx}
\usepackage{bm}
\usepackage{color}
\usepackage{hyperref}
 \usepackage{geometry} 
\begin{document}

\newcommand{\gl}{{\rm gl}}
\newcommand{\U}{{\rm U}}
\newcommand{\un}{{\rm u}}
\newcommand{\UOSp}{{\rm UOSp}}
\newcommand{\USp}{{\rm USp}}
\newcommand{\Herm}{{\rm Herm_{\odot}}}
\newcommand{\Sdet}{{\rm Sdet\,}}
\newcommand{\Str}{{\rm Str\,}}
\newcommand{\tr}{{\rm tr\,}}
\newcommand{\diag}{{\rm diag\,}}
\newcommand{\Id}{\mathbf{1}}
\newcommand{\CBE}{{\rm C}\beta{\rm E}}
\newcommand{\CtBE}{{\rm C}\widetilde{\beta}{\rm E}}
\newcommand{\GBE}{\chi{\rm G}\beta{\rm E}}
\newcommand{\LBE}{{\mathcal L}\beta{\rm E}}
\newcommand{\JBE}{{\rm J}\beta{\rm E}}

\title{Supersymmetry for  Products of Random Matrices
\thanks{Presented at ``Random Matrix Theory: Foundations and Applications" in Cracow, July 1-6 2014}
}

\author{Mario Kieburg
\address{Faculty of Physics, Bielefeld University, Postfach 100131, D-33501 Bielefeld, Germany}, {\it mkieburg@physik.uni-bielefeld.de}
}

\maketitle

\begin{abstract}
We consider the singular value statistics of products of independent random matrices. In particular we compute the corresponding averages of products of characteristic polynomials. To this aim we apply the projection formula recently introduced for chiral random matrix ensembles which serves as a short cut of the supersymmetry method.  The projection formula enables us to study the local statistics where free probability currently fails. To illustrate the projection formula and underline the power of our approach we calculate the hard edge scaling limit of the Meijer G-ensembles comprising the Wishart-Laguerre (chiral Gaussian), the Jacobi (truncated orthogonal, unitary or unitary symplectic) and the Cauchy-Lorentz (heavy tail) random matrix ensembles. All calculations are done for real, complex, and quaternion matrices in a unifying way. In the case of real and quaternion matrices the results are completely new and point to the universality of the hard edge scaling limit  for a product of these matrices, too. Moreover we identify the non-linear $\sigma$-models corresponding to product matrices.
\end{abstract}
  
\section{Introduction}\label{Sec:intro}

Sums and products of random matrices are the simplest generalization of random matrix theory (RMT) to introduce some kind of dimension. Sums of random matrices can be understood as a convolution and regularly appear in the field of Dyson's Brownian motion~\cite{Dys-Brow}. Product matrices are versatile as well. Applications of them can be found in mesoscopic physics~\cite{meso,been}, QCD~\cite{Osborn}, and wireless telecommunication~\cite{tele,AIKW13}. In the past years a lot of progress was made on products of random matrices, see the new review~\cite{AI15} reporting on this progress. For example, free probability has proven as an efficient tool for calculating the macroscopic level density~\cite{freeprob}. With the help of orthogonal polynomials one could calculate algebraic structures like determinants and Pfaffians, their kernels, and certain universal statistics on the  local scale of the spectrum~\cite{AIKW13,Meijer-G,Meijer-G-sing-bet-2,IK14,KKS15}. In particular products of random matrices drawn from Meijer G-ensembles (the weight is essentially given by Meijer G-functions, see \cite{ASbook} for a definition of these functions) exhibit a new kind of universal kernel in the hard scaling limit (microscopic limit around the origin). This limit is called Meijer G-kernel. Its name is reminiscent to the fact that the kernel essentially depends on Meijer G-functions. The ``standard candles" of RMT, the  Wishart-Laguerre ensemble~\cite{Wishart} ($\GBE$), the Cauchy-Lorentz ensemble~\cite{Heavy-tail} ($\LBE$), and the Jacobi ensemble~\cite{trunc} ($\JBE$) are particular cases of Meijer G-ensembles. Also products of matrices drawn from these three ensembles are Meijer G-ensembles since this class of ensembles is expected to be closed under matrix products.

Most results on the singular value statistics about product matrices are known for complex matrices ($\beta=2$), only. The only exception, the macroscopic level density, can be computed for real ($\beta=1$) and quaternion ($\beta=4$) matrices with  free probability~\cite{freeprob} because they share the level density with $\beta=2$. However the local statistics of the singular values is still highly involved for $\beta=1,4$ due to unknown group integrals like the Itzykson-Zuber integral~\cite{IZ80} and its polynomial counterpart~\cite{group-int,KKS15}. The projection formula recently proposed~\cite{KKG14} circumvents such problems. This formula is a short cut of the supersymmetry method~\cite{SUSY-method,superbosonization} and directly relates the original probability density with the weight in the dual superspace.

After introducing the required notation in Sec.~\ref{Sec:prelim} we briefly review the projection formula in Sec.~\ref{Sec:idea}. Thereby we only consider the average of a product of characteristic polynomials to keep the calculation simple. We emphasize that the projection formula holds for all three Dyson indices $\beta=1,2,4$ which is the strength of this approach.

In Sec.~\ref{Sec:standard}, we demonstrate via the three ensembles, $\GBE$, $\LBE$, and $\JBE$, how the projection formula works. Thereby we explicitly compute the well-known orthogonal polynomials for $\beta=2$ and show that the average of one characteristic polynomial for $\beta=1$ and the square root of a characteristic polynomial for $\beta=4$ is apart from some shifts in the parameters the same as in the case $\beta=2$. Another example is presented in Sec.~\ref{Sec:products} where we generalize the approach to a product of independently distributed matrices. Also for product matrices we explicitly calculate the orthogonal polynomials in the case $\beta=2$. However the completely new results are the ones for $\beta=1,4$ which are expressed in terms of integrals over Dyson's circular ensembles ($\CBE$) \cite{Dys-Circ}. In this way we show in Sec.~\ref{Sec:hard} that the universality in the hard edge scaling limit holds for real and quaternion product matrices, too. We are also able to identify the non-linear $\sigma$-models which are necessary when comparing the universal results with physical field theories.

\section{Preliminaries}\label{Sec:prelim}

We consider rectangular random matrices which are either real ($\beta=1$), complex ($\beta=2$), or quaternion ($\beta=4$). We are particularly interested in the singular value statistics of a random matrix
\begin{equation}\label{W-def}
 W\in\gl^{(\beta)}(n;n+\nu)=\left\{\begin{array}{cl} \mathbb{R}^{n\times(n+\nu)}, & \beta=1,\\ \mathbb{C}^{n\times(n+\nu)}, & \beta=2,\\ \mathbb{H}^{n\times(n+\nu)}, & \beta=4 \end{array}\right.
\end{equation}
distributed by $P(WW^\dagger)$. We assume $\nu=0$ in the following to keep the computations simple such that we choose the abbreviation $\gl^{(\beta)}(n)=\gl^{(\beta)}(n;n)$. Nonetheless this restriction is not that strong since a product of rectangular matrices can be always rephrased to a product of square matrices \cite{IK14}. Examples of such induced measures resulting from rectangular matrices are given in Sec.~\ref{Sec:products}.

Since we choose the complex representation of the quaternion numbers $\mathbb{H}$ in terms of Pauli matrices we introduce the convenient parameters
\begin{equation}\label{gamma-def}
\widetilde{\beta}=\frac{4}{\beta},\ \gamma=\left\{\begin{array}{cl} 1, & \beta=1,2, \\ 2, & \beta=4, \end{array}\right.\ \widetilde{\gamma}=\left\{\begin{array}{cl} 2, & \beta=1, \\ 1, & \beta=2,4. \end{array}\right.
\end{equation}
For the sake of readability  we restrict ourselves to partition functions of the form
\begin{equation}\label{part-def}
 Z(M)=\int d[W] P(WW^\dagger) {\det}^{1/(\gamma\widetilde{\gamma})}(WW^\dagger\otimes\Id_{\widetilde{\gamma}k}-M).
\end{equation}
The fixed matrix $M=\{M_{ab,ij}\}$ has the dimension $(\gamma n\times\gamma n)\otimes(\widetilde{\gamma}k\times \widetilde{\gamma}k)=\gamma\widetilde{\gamma} nk\times\gamma\widetilde{\gamma} nk$. It has to satisfy the symmetry
\begin{equation}\label{trans-def}
 M^T=\left\{\begin{array}{cc} \Id_n\otimes[\tau_2\otimes\Id_k]\,M\,\Id_n\otimes[\tau_2\otimes\Id_k], & \beta=1, \\ {[}\tau_2\otimes\Id_n]\otimes\Id_k\, M\,[\tau_2\otimes\Id_n]\otimes\Id_k, & \beta=4, \end{array}\right.
\end{equation}
where $\tau_2$ is the second Pauli matrix. Other properties of $M$ are not required.

The partition function~\eqref{part-def} needs an explanation. The determinant acts on the tensor space of $(\gamma n\times\gamma n)$ matrices containing the matrix $WW^\dagger$ and a space of dimension $(\widetilde{\gamma}k\times \widetilde{\gamma}k)$. In the case that $M=\Id_{\gamma n}\otimes\diag(m_1,\ldots,m_{\widetilde{\gamma}k})$  the determinant is a short hand notation for a product of characteristic polynomials of $WW^\dagger$ which is a well-known partition function in random matrix theory \cite{Gernotbook}. The reason why we wrote this product in such an uncommon,  compact form is the application we aim at, namely the singular value statistics of matrix products. Then the matrix $M$ does not take such a simple form.

Another particularity of Eq.~\eqref{part-def} which needs an explanation is the exponent of the determinant, $-1/(\gamma\widetilde{\gamma})$ and the matrix dimensions. In the case of complex matrices ($\beta=2$), the exponent and the dimensions become self-explanatory since they  become trivial, e.g. $-1/(\gamma\widetilde{\gamma})|_{\beta=2}=-1$.  When $W$ is real ($\beta=1$) then $WW^\dagger$ is real symmetric and $n\times n$ dimensional. The space dual to the polynomials consists of self-dual matrices. The resulting Kramers degeneracy cancels the exponent $1/2$ and doubles the dimension, $k\to2k$. Exactly the opposite happens in the case of a quaternion matrix $W$ ($\beta=4$). Due to its quaternion structure the dimension is doubled, $n\to2n$. However the dual space consists of symmetric matrices. Since symmetric matrices may have also odd dimensions we do not need a doubling of the dimension $k$. The corresponding square roots of the characteristic polynomials are exact and, thus, a polynomial because the spectrum of $WW^\dagger$ is Kramers degenerate. Such a square root is known as quaternion determinant and is equivalent to a Pfaffian determinant~\cite{Mehtabook}.

An important ingredient needed for the supersymmetry method is the invariance of the probability density $P$  under the transformation $P(WW^\dagger)=P(U WW^\dagger U^\dagger)$ for all $U\in\U^{(\beta)}(n)$ where
\begin{eqnarray}\label{group-def}
\U^{(\beta)}(n)=\left\{ \begin{array}{cl} {\rm O}(n), & \beta=1,\\ {\rm U}(n), & \beta=2,\\ {\rm USp}(2n), & \beta=4. \end{array}\right.
\end{eqnarray}
Only due to this invariance it is possible to find an integral over a supermatrix whose dimension is independent of the ordinary dimension $n$ and which yields exactly the same partition function as Eq.~\eqref{part-def}. This can be achieved in four steps which we briefly sketch in section~\ref{Sec:idea}.

For this purpose we have to introduce two  supermatrix spaces and one ordinary matrix space. Let $p,q,N\in\mathbb{N}$, and $\U(p|q)$ and $\UOSp(p|2q)$ be the unitary and the unitary ortho-symplectic supergroup, respectively, see \cite{Zirn,supergroups,KKG09}. The space of rectangular supermatrices is defined by
\begin{equation}\label{SUSY-rec-def}
\gl^{(\beta)}(p|q;p'|q')=\un^{(\beta)}(p+p'|q+q')/[\un^{(\beta)}(p|q)\times\un^{(\beta)}(p'|q')],
\end{equation}
where $\un^{(\beta)}(p|q)$ is the Lie  superalgebra of the supergroup
\begin{equation}\label{SUSY-group-def}
\U^{(\beta)}(p|q)=\left\{\begin{array}{cl} \UOSp^{(+)}(p|2q), & \beta=1,\\ \U(p|q), & \beta=2,\\ \UOSp^{(-)}(2p|q), & \beta=4.\end{array}\right.
\end{equation}
The coset is taken via the addition as a group action on the Lie  superalgebra. Therefore a matrix $\rho\in\gl^{(\beta)}(p|q;N)$ is $(\gamma p|\widetilde{\gamma} q)\times (\gamma p|\widetilde{\gamma} q)$ dimensional and has the following form
\begin{equation}\label{SUSY-matrix-def}
\rho=\left[\begin{array}{cc} \rho_{\rm BB} & \rho_{\rm BF} \\ \rho_{\rm FB} & \rho_{\rm FF} \end{array}\right].
\end{equation}
The $\gamma p\times \gamma p$ dimensional boson-boson block $\rho_{\rm BB}$ and the $\widetilde{\gamma} q\times \widetilde{\gamma} q$ dimensional fermion-fermion block $\rho_{\rm FF}$ comprise commuting variables while the other two block contain anti-commuting ones. 

We employ the same notation for the two inequivalent fundamental representations of the supergroup $\UOSp(p|2q)$ as in \cite{supergroups,KKG09} where the superscripts indicate the transformation property under the complex conjugation, i.e.
\begin{equation}\label{SUSY-real-def}
 \rho^*=\left\{\begin{array}{cc} \diag(\Id_{p},-\imath\tau_2\otimes\Id_q)\,\rho\,\diag(\Id_{p'},\imath\tau_2\otimes\Id_{q'}), & \beta=1, \\ \diag(-\imath\tau_2\otimes\Id_{p},\Id_{q})\,\rho\,\diag(\imath \tau_2\otimes\Id_{p'},\Id_{q'}), & \beta=4 \end{array}\right.
\end{equation}
for $\rho\in\gl^{(\beta)}(p|q;p'|q')$ and 
\begin{equation}\label{SUSY-real-group-def}
 U^*=\left\{\begin{array}{cc} \diag(\Id_{p},-\imath\tau_2\otimes\Id_q)\,U\,\diag(\Id_{p},\imath\tau_2\otimes\Id_q), & \beta=1, \\ \diag(-\imath\tau_2\otimes\Id_{p},\Id_{q})\,U\,\diag(\imath\tau_2\otimes\Id_{p},\Id_{q}), & \beta=4 \end{array}\right.
\end{equation}
for $U\in\U^{(\beta)}(p|q)\subset\U(\gamma p|\widetilde{\gamma} q)$. The two relations~\eqref{SUSY-real-def} and \eqref{SUSY-real-group-def} are generalization of the definitions of real and quaternion matrices to superspace. 

The ordinary matrix space announced is the coset
\begin{equation}\label{C-beta-E}
 \CBE(\gamma k)=\left\{\begin{array}{cl} \U(k)/{\rm O}(k), & \beta=1,\\ {[\U(k)\times\U(k)]}/\U(k)\simeq\U(k), & \beta=2,\\ \U(2k)/\USp(2k), & \beta=4 \end{array}\right.
\end{equation}
equipped with a normalized Haar measure $d\mu(U)$ induced by the Haar measures on the defining groups. These three sets are the circular ensembles first studied by Dyson~\cite{Dys-Circ}. These co-sets are also the fermionic part of the supermatrices involved in the superbosonization formula~\cite{superbosonization}. Since we only discuss the average of products of determinants and not ratios superbosonization reduces to bosonization only involving the circular ensembles~\eqref{C-beta-E}. Let us recall the properties of a matrix $U\in\CBE(\gamma k)$. The matrix $U$ is unitary and satisfies the symmetries $U^T=U$ for $\beta=1$ and $U^T=(\tau_2\otimes\Id_k)U(\tau_2\otimes\Id_k)$ for $\beta=4$.

Also the superdeterminant and the supertrace play an important role in the ensuing calculations. They are defined via the ordinary determinant and trace and explicitly read
\begin{equation}\label{SUSY-def}
 \Sdet\rho=\frac{\det(\rho_{\rm BB}-\rho_{\rm BF}\rho_{\rm FF}^{-1}\rho_{\rm FB})}{\det\rho_{\rm FF}},\ \Str\rho=\tr\rho_{\rm BB}-\tr\rho_{\rm FF}
\end{equation}
 for an arbitrary square supermatrix $\rho\in\gl^{(\beta)}(p|q;p|q)$
whose fermion-fermion block $\rho_{FF}$ is invertible.
The definitions are chosen in such a way that many properties of the trace and the determinant carry over to superspace. For example the circularity $\Str AB=\Str BA$, the factorization $\Sdet AB=\Sdet A\, \Sdet B$, and the relation $\ln\Sdet A=\Str\ln A$ still hold for two arbitrary invertible square supermatrices $A$ and $B$. The circularity property of the supertrace works for rectangular supermatrices, as well. A more profound introduction in supersymmetric analysis and algebra can be found in \cite{SUSY}.

\section{What is the Projection Formula?}\label{Sec:idea}

The  projection formula in its general form projects functions living on a very large superspace to functions on a much smaller superspace \cite{KKG14}. In this way it directly relates the original weight $P$ to a weight $Q$ in the smaller superspace. Hence the projection formula is a short cut of the supersymmetry method~\cite{KKG14}. For our particular purposes the large superspace is $\gl^{(\beta)}(n+\widetilde{\gamma}l|\gamma l;n|0)$ with $l$ being an integer larger than or equal to $k/\gamma$. The enlargement of the dimensions $k\to 2l$ in the case $k$ odd and $\beta=4$ is crucial. The reason is a Cauchy-like integration theorem~\cite{Cauchy-int,KKG09} first derived in a general framework by Wegner~\cite{Wegner} which only applies to an even dimensional reduction of a matrix space in the case of $\beta=1,4$.

In the first step of deriving the projection formula we need the following version of this Cauchy-like theorem~\cite{KKG14}
\begin{equation}\label{Cauchy}
 P(WW^\dagger)=\frac{\int d[\widehat{\Omega}] P(\Omega\Omega^\dagger)}{\int d[\widehat{\Omega}] \exp[-\Str\widehat{\Omega}\widehat{\Omega}^\dagger]}
\end{equation}
with $W\in\gl^{(\beta)}(n)=\gl^{(\beta)}(n|0;n|0)$ and $\widehat{\Omega}\in\gl^{(\beta)}(\widetilde{\gamma}l|\gamma l;n|0)$. The matrices are embedded as follows
\begin{equation}\label{split}
 \Omega=\left[\begin{array}{c} W \\ \widehat{\Omega}\end{array}\right]=\left[\begin{array}{c} W' \\ \Omega'\end{array}\right]\in\gl^{(\beta)}(n+\widetilde{\gamma}l|\gamma l;n).
\end{equation}
The second splitting in $W'\in\gl^{(\beta)}(n+\widetilde{\gamma}l|\gamma l-k;n|0)$ and $\Omega'\in\gl^{(\beta)}(0|k;n|0)$ becomes relevant in the third step of the derivation of the projection formula. The measure $d[\widehat{\Omega}]$ is the product of all differentials of independent matrix entries of $\widehat{\Omega}$. The normalization with a Gaussian is true because the proportionality constant is independent of $P$ and thus can be fixed by any weight.

In Eq.~\eqref{Cauchy} we have chosen a supersymmetric extension of $P$ to the superspace $\gl^{(\beta)}(n+\widetilde{\gamma}l|\gamma l;n|0)$ which is by far not unique. However the final result is independent of this choice as already discussed in \cite{IK14}. Such an extension indeed exists for a smooth distribution $P$. Since $P$ is invariant under the group $\U^{(\beta)}(n)$ we can apply the Cayley-Hamilton theorem implying that $P$ can be expressed in matrix invariants like traces and determinants of $WW^\dagger$. Those invariants have  invariant extensions, namely the supertrace and the superdeterminant, cf. Eq.~\eqref{SUSY-def}. 

In the next step we rewrite the determinant in Eq.~\eqref{part-def} as a Gaussian integral over a matrix $V=\{V_{aj}\}\in\gl^{(\beta)}(0|k;n|0)$ which only consists of Grassmann (anti-commuting) variables~\cite{SUSY},
\begin{eqnarray}\label{Gauss}
&&{\det}^{1/(\gamma\widetilde{\gamma})}(WW^\dagger\otimes\Id_{\widetilde{\gamma}k}-M)\\
&=&\frac{\int d[V] \exp[\tr VWW^\dagger V^\dagger-\sum_{a,b=1}^{\widetilde{\gamma}k}\sum_{i,j=1}^{\gamma n}M_{ab,ij}V_{ai}V_{bj}^*]}{\int d[V] \exp[\tr VV^\dagger]}.\nonumber
\end{eqnarray}
Then the partition function is up to a constant
\begin{equation}\label{part-1}
 Z(M)\propto\int d[\Omega]d[V]  P(\Omega\Omega^\dagger) \exp\left[-\Str \Omega\Omega^\dagger \widehat{V}^\dagger\widehat{V}-\sum_{a,b=1}^{\widetilde{\gamma}k}\sum_{i,j=1}^{\gamma n}M_{ab,ij}V_{ai}V_{bj}^*\right]
\end{equation}
with
\begin{equation}\label{V-def}
\widehat{V}=\left[\begin{array}{cc} 0 & 0 \\ V & 0\end{array}\right],\ \widehat{V}^\dagger=\left[\begin{array}{cc} 0 & V^\dagger \\ 0 & 0\end{array}\right]\in\gl^{(\beta)}(n+\widetilde{\gamma}l|\gamma l;n+\widetilde{\gamma}l|\gamma l).
\end{equation}
The first $(\gamma n+2\gamma\widetilde{\gamma} l-\widetilde{\gamma}k)$ rows and the last $2\gamma\widetilde{\gamma}l$ columns of $\widehat{V}$ are equal to $0$. The change of the sign in front of the first term in the exponential function relates to the fact that Grassmann variables are anti-commuting.

The integrals over $V$ and $\Omega$ can be interchanged such that we find the function
\begin{equation}\label{Four-1}
\widehat{P}(\widehat{V}^\dagger\widehat{V})=\int d[\Omega]P(\Omega\Omega^\dagger)\exp[-\Str \Omega\Omega^\dagger \widehat{V}^\dagger\widehat{V}].
\end{equation}
The invariance of $P(\Omega\Omega^\dagger)=P(U\Omega\Omega^\dagger U^\dagger)$ for all $U\in\U^{(\beta)}(n+\widetilde{\gamma}l|\gamma l)$ carries over to a symmetry for $\widehat{P}(\widehat{V}^\dagger\widehat{V})=\widehat{P}(U\widehat{V}^\dagger\widehat{V}U^\dagger)$  for all $U\in\U^{(\beta)}(n+\widetilde{\gamma}l|\gamma l)$. Therefore the following duality holds
\begin{equation}\label{duality}
\widehat{P}(\widehat{V}^\dagger\widehat{V})=\widehat{P}(\widehat{V}\widehat{V}^\dagger)
\end{equation}
which is the third important step of the derivation. Employing the definition~\eqref{Four-1} backwards the partition function is
\begin{eqnarray}
 Z(M)&\propto&\int d[\Omega]d[V]  P(\Omega\Omega^\dagger) \exp\left[\tr V^\dagger\Omega'\Omega^{\prime\,\dagger} V-\sum_{a,b=1}^{\widetilde{\gamma}k}\sum_{i,j=1}^{\gamma n}M_{ab,ij}V_{ai}V_{bj}^*\right]\nonumber\\
 &\propto&\int d[\Omega']Q(\Omega'\Omega^{\prime\,\dagger}){\det}^{1/(\gamma\widetilde{\gamma})}(\Id_{\gamma n}\otimes\Omega'\Omega^{\prime\,\dagger}-M).\label{part-2}
\end{eqnarray}
In the last step we integrated over the remaining degrees of freedom $W'$, cf. the splitting~\eqref{split}, which do not show up in the determinant. This integration yields the function
\begin{equation}\label{Q-def-a}
 Q(\Omega'\Omega^{\prime\,\dagger})\propto\int d[W']P\left(\left[\begin{array}{cc} W'W^{\prime\,\dagger} & W'\Omega^{\prime\,\dagger} \\ \Omega'W^{\prime\,\dagger} & \Omega'\Omega^{\prime\,\dagger}\end{array}\right]\right) .
\end{equation}
This equation is the essence of the projection formula. The remaining things to do is cosmetics.

We want to  express the dyadic matrix $\Omega'\Omega^{\prime\,\dagger}$ as a single square matrix $U$ which is an element in $\CtBE(\gamma k)$. Note that the circular ensemble really relates to the Dyson index $\widetilde{\beta}=4/\beta$ and not $\beta$ which originates from the symmetries fulfilled by $V$.

Exactly this is done in the last step. We apply the superbosonization formula~\cite{superbosonization} which reduces to pure bosonization in our case.  This yields the partition function
\begin{equation}
 Z(M)=\int d\mu(U)Q(U){\det}^{1/(\gamma\widetilde{\gamma})}(\Id_{\gamma n}\otimes U-M) {\det}^{-n/\widetilde{\gamma}}U.\label{part-3}
\end{equation}
with the normalized distribution
\begin{equation}\label{Q-def-b}
 Q(U)=\frac{\int d[W_1]d[W_2]P\left(\left[\begin{array}{cc} W_1W_1^\dagger+W_2W_2^\dagger & W_2U^{1/2} \\ U^{1/2}W_2^\dagger & U \end{array}\right]\right)}{\int d\mu(U) d[W_1]d[W_2]\det^{-n/\widetilde{\gamma}}U\exp[-\Str (W_1W_1^\dagger+W_2W_2^\dagger)+\tr U]}.
\end{equation}
The reduction of the integral~\eqref{Q-def-a} to the final expression~\eqref{Q-def-b} as an integral over the two matrices $W_1\in\gl^{(\beta)}(n+\widetilde{\gamma}l | \gamma l-k; n+\widetilde{\gamma}l | \gamma l-k)$ and $W_2\in\gl^{(\beta)}(n+\widetilde{\gamma}l | \gamma l-k;0|k)$  was done in \cite{KKG09} and is skipped here due to the lack of space.

 We remark that apart from the case $k$ odd and $\beta=4$ the auxiliary parameter $l$ can be chosen $l=k/\gamma$. Then the matrix $W_1$ is an ordinary square matrix and $W_2$ is a rectangular matrix only consisting of Grassmann variables.

\section{Application to Standard Random Matrix Ensembles}\label{Sec:standard}

Three particular cases of Meijer G-ensembles are the Gaussian $\GBE$, the heavy-tailed $\LBE$, and the compactly supported $\JBE$. We discuss them in subsections~\ref{Sec:Lag}, \ref{Sec:Lor}, and \ref{Sec:Jac}, respectively. These ensembles play important roles in a vast of applications and cover a broad range of systems~\cite{Wishart,Meijer-G,Meijer-G-sing-bet-2,Gernotbook,trunc,Heavy-tail}.

\subsection{Wishart-Laguerre (Gaussian) Ensemble}\label{Sec:Lag}

The first ensemble we consider is the $\GBE$,
\begin{eqnarray}\label{W-L-def}
P_{\rm WL}(WW^\dagger)\propto{\det}^{\nu/\widetilde{\gamma}}WW^\dagger \exp[-\tr WW^\dagger/\Gamma^2]
\end{eqnarray}
with $\nu\in\mathbb{N}_0$ and $\Gamma>0$. It is the oldest random matrix ensemble first studied by Wishart~\cite{Wishart}. The determinant in front of the Gaussian originates from a transformation of a rectangular matrix $W'\in\gl^{(\beta)}(n,n+\nu)$ to the square matrix $W\in\gl^{(\beta)}(n)$. Therefore one can understand Eq.~\eqref{W-L-def} as an induced measure~\cite{IK14}. The corresponding weight $Q_{\rm WL}$ is given by Eq.~\eqref{Q-def-b},
\begin{eqnarray}
 Q_{\rm WL}(U)&\propto&\int d[W_1]d[W_2] \Sdet^{\nu/\widetilde{\gamma}}\left[\begin{array}{cc} W_1W_1^\dagger+W_2W_2^\dagger & W_2U^{1/2} \\ U^{1/2}W_2^\dagger & U \end{array}\right]\nonumber\\
 &&\times\exp[-\Str(W_1W_1^\dagger+W_2W_2^\dagger)+\tr U/\Gamma^2]\nonumber\\
 &\propto&{\det}^{-\nu/\widetilde{\gamma}}Ue^{\tr U/\Gamma^2}.\label{Q-W-L}
\end{eqnarray}
Therefore the partition function~\eqref{part-def} for $P_{\rm WL}(WW^\dagger)$ reads
\begin{equation}
 Z_{\rm WL}(M)=\frac{\int d\mu(U) {\det}^{-(n+\nu)/\widetilde{\gamma}}U{\det}^{1/(\gamma\widetilde{\gamma})}(\Id_{\gamma n}\otimes U-M)e^{\tr U/\Gamma^2}}{\int d\mu(U) {\det}^{-(n+\nu)/\widetilde{\gamma}}U e^{\tr U/\Gamma^2}}.\label{part-W-L}
\end{equation}
This result agrees with the one derived in \cite{KKG09}. The normalization can be fixed by considering the expansion of the partition function for large $M$.

The result~\eqref{part-W-L} exhibits nice implications. For example the case $k=\gamma$ and $M=m\Id_{\gamma^2\widetilde{\gamma} n}$ is equal to the orthogonal polynomials for $\beta=2$ and to the skew-orthogonal polynomials of even order for $\beta=1,4$, see \cite{Mehtabook}. Hence the contour for $\beta=2$ is a representation of the modified Laguerre polynomials $L_n^{(\nu)}$, see  \cite{ASbook}, i.e.
\begin{eqnarray}
 Z_{\rm WL}^{(\beta=2,k=1)}(m\Id_n)&\propto&\oint dz z^{-(n+\nu+1)}(z-m)^ne^{ z/\Gamma^2}\nonumber\\
 &\propto&\sum_{j=0}^{n}\frac{1}{j!(n-j)!(\nu+j)!}\left(-\frac{m}{\Gamma^2}\right)^j\propto L_n^{(\nu)}\left(\frac{m}{\Gamma^2}\right).\label{pol-W-L}
\end{eqnarray}
These polynomials also appear for $\beta=1,4$ if we set $k=1$. Only the argument $m$ is modified to $\widetilde{\gamma}m$. Interestingly, the case $\beta=4$ is an average over a square root of a determinant which is equivalent to a Pfaffian.

For the case $k=2\gamma$ and $M=\Id_{\gamma^2\widetilde{\gamma} n}\otimes\diag(m_1,m_2)$ we find one of the kernels corresponding to the $\GBE$ \cite{Mehtabook}. When computing the contour integral~\eqref{part-W-L} we immediately find the corresponding Christoffel-Darboux formulas.

\subsection{Cauchy-Lorentz Ensemble}\label{Sec:Lor}

The $\LBE$ is the next case we want to study. It is defined by the probability density~\cite{Heavy-tail,KKG14}
\begin{eqnarray}\label{C-L-def}
P_{\rm CL}(WW^\dagger)\propto{\det}^{\nu/\widetilde{\gamma}}WW^\dagger {\det}^{-\mu}(\Gamma^2\Id_{\gamma n}+WW^\dagger)
\end{eqnarray}
with $\Gamma>1$, $\nu\in\mathbb{N}_0$ and $\mu>k/\gamma+(2n+\nu)/\widetilde{\gamma}-(\gamma\widetilde{\gamma}-1)/2$ for guaranteeing the convergence of the integral~\eqref{part-def}\footnote{Note that the inequality satisfied by $\mu$ in \cite{KKG14} contains a mistake which we have corrected here. The inequality can be found by performing a singular value decomposition of $W$ and then reading off the algebraic behaviour at infinity.}. It is a heavy-tailed distribution and was employed for modelling financial correlations~\cite{Heavy-tail}.

The choice $\Gamma>1$ is convenient for the projection formula but is not a restriction at all because it only rescales the ensemble. The term ${\det}^{\nu/\widetilde{\gamma}}WW^\dagger$ can be again understood as a remnant of a rectangular matrix $W'\in\gl^{(\beta)}(n,n+\nu)$. However we underline that such a transformation from $W'$ to $W$ also changes the exponent $\mu$.

The weight for the dual space is calculated by Eq.~\eqref{Q-def-b},
\begin{eqnarray}
 Q_{\rm CL}(U)&\propto&\int d[W_1]d[W_2] \Sdet^{\nu/\widetilde{\gamma}}\left[\begin{array}{cc} W_1W_1^\dagger+W_2W_2^\dagger & W_2U^{1/2} \\ U^{1/2}W_2^\dagger & U \end{array}\right]\nonumber\\
 &&\hspace*{-0.3cm}\times\Sdet^{-\mu}\left[\begin{array}{cc} \Gamma^2\Id_{\gamma n+\gamma\widetilde{\gamma}l|\gamma\widetilde{\gamma}l-\widetilde{\gamma}k}+W_1W_1^\dagger+W_2W_2^\dagger & W_2U^{1/2} \\ U^{1/2}W_2^\dagger & \Gamma^2\Id_{\widetilde{\gamma}k}+U \end{array}\right]\nonumber\\
 &\propto&{\det}^{-\nu/\widetilde{\gamma}}U{\det}^{\mu}(\Gamma^2\Id_{\widetilde{\gamma}k}+U)\int d[W_1]d[W_2] \Sdet^{\nu/\widetilde{\gamma}}W_1W_1^\dagger\nonumber\\
 &&\hspace*{-0.3cm}\times\Sdet^{-\mu}\left[\Gamma^2\Id_{\gamma n+\gamma\widetilde{\gamma}l|\gamma\widetilde{\gamma}l-\widetilde{\gamma}k}+W_1W_1^\dagger+\Gamma^2W_2(\Gamma^2\Id_{\widetilde{\gamma}k}+U)^{-1}W_2^\dagger \right]\nonumber\\
 &\propto& {\det}^{-\nu/\widetilde{\gamma}}U{\det}^{\mu-k/\gamma-n/\widetilde{\gamma}}(\Gamma^2\Id_{\widetilde{\gamma}k}+U)\label{Q-C-L}
\end{eqnarray}
In the last step we have rescaled $W_2\to W_2(\Gamma^2\Id_{\widetilde{\gamma}k}+U)^{1/2}$ such that the remaining integrals are independent of $U$. Thereby we recall that the Berezinian (Jacobian in superspace) is ${\det}^{-k/\gamma-n/\widetilde{\gamma}}(\Gamma^2\Id_{\widetilde{\gamma}k}+U)$.  Hence we end up with the partition function
\begin{equation}
 Z_{\rm CL}(M)=\frac{\int d\mu(U) {\det}^{-\frac{n+\nu}{\widetilde{\gamma}}}U{\det}^{\mu-\frac{k}{\gamma}-\frac{n}{\widetilde{\gamma}}}(\Gamma^2\Id_{\widetilde{\gamma}k}+U){\det}^{\frac{1}{\gamma\widetilde{\gamma}}}(\Id_{\gamma n}\otimes U-M)}{\int d\mu(U) {\det}^{-(n+\nu)/\widetilde{\gamma}}U{\det}^{\mu-k/\gamma-n/\widetilde{\gamma}}(\Gamma^2\Id_{\widetilde{\gamma}k}+U)}.\label{part-C-L}
\end{equation}
Starting from this formula one can again easily deduce the orthogonal or skew-orthogonal polynomials, the kernel involving two characteristic polynomials, and the Christoffel-Darboux formula associated to this kernel. For example the orthogonal polynomials corresponding to the complex $\LBE={\rm LUE}$ is
\begin{eqnarray}
 Z_{\rm CL}^{(\beta=2,k=1)}(m\Id_n)&\propto&\oint dz z^{-(n+\nu+1)}(\Gamma^2+z)^{\mu-n-1}(z-m)^n\label{pol-CL}\\
 &\propto&\sum_{j=0}^{n}\frac{1}{j!(n-j)!(\nu+j)!\Gamma[\mu-n-\nu-j]}\left(-\frac{m}{\Gamma^2}\right)^j.\nonumber
\end{eqnarray}
This polynomial can be understood as a Jacobi polynomial when  analytically continuing the parameters to negative values, cf. Eq.~\eqref{pol-J}. The same polynomials pop up for $\beta=1,4$ when setting $k=1$. This time we have only to change the exponent $\mu\to\widetilde{\gamma}\mu-\widetilde{\gamma}/\gamma+1$.

\subsection{Jacobi (Truncated Unitary) Ensemble}\label{Sec:Jac}

The $\JBE$ is defined by~\cite{trunc}
\begin{eqnarray}\label{J-def}
P_{\rm J}(WW^\dagger)\propto{\det}^{\nu/\widetilde{\gamma}}WW^\dagger {\det}^{\kappa}(\Gamma^2\Id_{\gamma n}-WW^\dagger)\Theta(\Gamma^2\Id_{\gamma n}-WW^\dagger),
\end{eqnarray}
where $\nu\in\mathbb{N}_0$, $\kappa>-1/(2\gamma)$. The Heaviside step function for matrices  $\Theta$ is unity if the matrix is positive definite and otherwise vanishes. Again the scaling $\Gamma>1$ is only introduced to avoid problems with the contour integrals in the dual space. In the case $\gamma\widetilde{\gamma}\mu\in\mathbb{N}_0$ the random matrix $W$ distributed by Eq.~\eqref{J-def}  can be understood as a truncation of an orthogonal ($\beta=1$), a unitary ($\beta=2$), or a unitary symplectic ($\beta=4$) matrix, respectively, see \cite{trunc,IK14}.

To apply the projection formula we have first to find the supersymmetric generalization of the Heaviside step function. For this reason we write this function as $\Theta(\Gamma^2\Id_{\gamma n}-W^\dagger W)$. Then it is clear that this function reads in terms of the supermatrix $\Omega$ as $\Theta(\Gamma^2\Id_{\gamma n}-\Omega^\dagger \Omega)$ because the dyadic matrix $\Omega^\dagger \Omega$ has still an ordinary dimension and can be embedded in the space of $\gamma n\times \gamma n$ matrices by a Taylor expansion in the Grassmann valued matrix entries. Such a Taylor expansion is always finite since Grassmann variables are nilpotent. Hence we do not have to fear any problems of convergence.

Let $n,p,q\in\mathbb{N}$ and $V\in\gl^{(\beta)}(p|q;n|0)$. Then the extension of the Heaviside step function is done by a limit,
\begin{eqnarray}
\Theta(\Id_{\gamma n}-V^\dagger V)&=&\lim_{\epsilon\to\infty}{\det}^{-1}(\Id_{\gamma n}+e^{-\epsilon}e^{\epsilon V^\dagger V})\nonumber\\
&=&\lim_{\epsilon\to\infty}\exp\left[\sum_{j=1}^\infty\frac{(-1)^je^{-j\epsilon}}{j}\tr e^{j\epsilon V^\dagger V}\right].\label{SUSY-Theta-def-a}
\end{eqnarray}
This limit vanishes if one or more eigenvalues of the numerical part of the dyadic matrix $V^\dagger V$ is larger than $1$. We emphasize that indeed only the numerical part matters and not the nilpotent terms because of the Taylor expansion in the latter. In the next step we employ the duality $\tr e^{j\epsilon V^\dagger V}= \gamma (n-p)+\widetilde{\gamma} q+\Str e^{j\epsilon  VV^\dagger}$. We have
\begin{eqnarray}
\Theta(\Id_{\gamma n}-V^\dagger V)&=&\lim_{\epsilon\to\infty}(1+e^{-\epsilon})^{\gamma (p-n)-\widetilde{\gamma} q}\Sdet^{-1}(\Id_{\gamma p|\widetilde{\gamma} q}+e^{-\epsilon}e^{\epsilon V V^\dagger})\nonumber\\
&=&\lim_{\epsilon\to\infty}{\det}^{-1}(\Id_{\gamma p|\widetilde{\gamma} q}+e^{-\epsilon}\{e^{\epsilon V V^\dagger}\}_{\rm BB})\nonumber\\
&=&\Theta(\Id_{\gamma p}- \{VV^\dagger\}_{\rm BB}^{\rm num}) .\label{SUSY-Theta-def-b}
\end{eqnarray}
The Heaviside step function is only taken for the numerical part $\{VV^\dagger\}_{\rm BB}^{\rm num}$ of the boson-boson block of the dyadic matrix $VV^\dagger$. Any expansion in the nilpotent terms yields a polynomial in $\epsilon$ which are suppressed by the exponential $e^{-\epsilon}$. This implies that the other three blocks of the supermatrix $e^{\epsilon V V^\dagger}$ cannot contribute because they are polynomials in $\epsilon$. The boson-boson block is $\{e^{\epsilon V V^\dagger}\}_{\rm BB}=e^{\epsilon \{V V^\dagger\}_{\rm BB}^{\rm num}}(1+f(\epsilon))$ with $f$ a polynomial with $f(0)=0$. Therefore Eq.~\eqref{SUSY-Theta-def-b} is the correct generalization of the Heaviside step function to the superspace. Interestingly the Taylor expansion in the nilpotent terms have no influence on the Heaviside step function. But this behaviour has to be expected because the Taylor expansion can only have an effect on the boundary. Only there one or more eigenvalues of the numerical part $\{VV^\dagger\}_{\rm BB}^{\rm num}$ are equal to $1$ where the value of the function may change. However, the supersymmetric Heaviside step function vanishes at the boundary, too, due to the expansion in the nilpotent terms yielding an inverted polynomial  in $\epsilon$, e.g. ${\det}^{-1}(\Id_{\gamma p|\widetilde{\gamma} q}+e^{-\epsilon}\{e^{\epsilon V V^\dagger}\}_{\rm BB})\overset{\{V V^\dagger\}_{\rm BB}\to\Id_{\gamma p}}{\longrightarrow} 1/f(\epsilon)\overset{\epsilon\to\infty}{\to}0$ with $f$ a polynomial.

We employ Eq.~\eqref{SUSY-Theta-def-b} in our setting and recognize that the matrix $U$ is not a part of the boson-boson block of the matrix argument of $P_{\rm J}$ in Eq.~\eqref{Q-def-b}. Hence, the function in the dual space is
\begin{eqnarray}
 Q_{\rm J}(U)&\propto&\int d[W_1] d[W_2] \Sdet^{\nu/\widetilde{\gamma}}\left[\begin{array}{cc} W_1W_1^\dagger+W_2W_2^\dagger & W_2U^{1/2} \\ U^{1/2}W_2^\dagger & U \end{array}\right]\nonumber\\
 &&\times\Sdet^{\kappa}\left[\begin{array}{cc} \Gamma^2\Id_{\gamma n+\gamma\widetilde{\gamma}l|\gamma\widetilde{\gamma}l-\widetilde{\gamma}k}-W_1W_1^\dagger-W_2W_2^\dagger & W_2U^{1/2} \\ U^{1/2}W_2^\dagger & \Gamma^2\Id_{\widetilde{\gamma}k}-U \end{array}\right]\nonumber\\
 &&\times\Theta(\Gamma^2\Id_{\gamma n+\gamma\widetilde{\gamma}l}-\{W_1W_1^\dagger+W_2W_2^\dagger\}_{\rm BB}^{\rm num})\nonumber\\
 &\propto& {\det}^{-\nu/\widetilde{\gamma}}U{\det}^{-\kappa-k/\gamma-n/\widetilde{\gamma}}(\Gamma^2\Id_{\widetilde{\gamma}k}-U).\label{Q-J}
\end{eqnarray}
We underline that the boson-boson block of $W_2W_2^\dagger$ only consists of nilpotent parts such that it does not contribute to the Heaviside step function.  The corresponding partition function is
\begin{equation}
 Z_{\rm J}(M)=\frac{\int d\mu(U) {\det}^{-\frac{n+\nu}{\widetilde{\gamma}}}U{\det}^{-\kappa-\frac{k}{\gamma}-\frac{n}{\widetilde{\gamma}}}(\Gamma^2\Id_{\widetilde{\gamma}k}-U){\det}^{\frac{1}{\gamma\widetilde{\gamma}}}(\Id_{\gamma n}\otimes U-M)}{\int d\mu(U) {\det}^{-(n+\nu)/\widetilde{\gamma}}U{\det}^{-\kappa-k/\gamma-n/\widetilde{\gamma}}(\Gamma^2\Id_{\widetilde{\gamma}k}-U)}.\label{part-J}
\end{equation}
One can readily check the correctness of this result by calculating the orthogonal or skew-orthogonal polynomials and the kernel involving two characteristic polynomials. For example, with the help of the residue theorem we generate the polynomials,
\begin{eqnarray}
 Z_{\rm J}^{(\beta=2,k=1)}(m\Id_n)\hspace*{-0.1cm}&\propto&\oint dz z^{-(n+\nu+1)}(\Gamma^2-z)^{-(n+\kappa+1)}(z-m)^n\label{pol-J}\\
 &\propto&\sum_{j=0}^{n}\frac{\Gamma[n+\kappa+\nu+j+1]}{j!(n-j)!(\nu+j)!}\left(-\frac{m}{\Gamma^2}\right)^j\propto P_n^{(\kappa,\nu)}\left(\frac{2m}{\Gamma^2}-1\right) ,\nonumber
\end{eqnarray}
where $P_n^{(\kappa,\nu)}$ are the Jacobi polynomials with respect to the weight $(1-x)^\kappa(1+x)^\nu\Theta(1-x^2)$, see~\cite{ASbook}. As in the case of the $\LBE$ we find the same polynomials for $\beta=1,4$ and $k=1$ when replacing the exponent $\kappa\to\widetilde{\gamma}\kappa+\widetilde{\gamma}/\gamma-1$.

We also obtain the well-known Christoffel-Darboux formula of the Jacobi polynomials by setting  $k=2$, $\beta=2$, and $M=\Id_n\otimes\diag(m_1,m_2)$. Then the integral reduces to a double contour integral after diagonalizing $U$.

\section{Application to Product Matrices}\label{Sec:products}

The computation of the partition function for a product of $L$ matrices $W\to W^{(L)}=\prod_{j=1}^LW_j=W_1\cdots W_L$ independently distributed by $P(WW^\dagger)\to\prod_{j=1}^LP_j(W_jW_j^\dagger)$ works in a similar way as for a single matrix. Starting from the partition function
\begin{eqnarray}
 Z_{\Pi}(M)&=&\hspace*{-0.15cm}\int\left(\prod_{j=1}^L d[W_j] P_j(W_jW_j^\dagger)\right) {\det}^{1/(\gamma\widetilde{\gamma})}\left[W^{(L)}\left(W^{(L)}\right)^\dagger\otimes\Id_{\widetilde{\gamma}k}-M\right]\nonumber\\
 &=&\hspace*{-0.15cm}\int\left(\prod_{j=1}^L d[W_j] P_j(W_jW_j^\dagger)\right){\det}^{k/\gamma}W^{(L-1)}\left(W^{(L-1)}\right)^\dagger\label{part-prod-a}\\
 &&\times {\det}^{1/(\gamma\widetilde{\gamma})}\left[W_L W_L^\dagger\otimes\Id_{\widetilde{\gamma}k}-X_{L-1}^{-1}M Y_{L-1}^{-1}\right]\nonumber
\end{eqnarray}
with $X_{L-1}=W^{(L-1)}\otimes\Id_{\widetilde{\gamma}k}$ and $Y_{L-1}=(W^{(L-1)})^\dagger\otimes\Id_{\widetilde{\gamma}k}$, we apply the projection formula for $W_L$ after replacing the matrix $M\to X_{L-1}^{-1}M Y_{L-1}^{-1}$. Then we obtain
\begin{eqnarray}
 Z_{\Pi}(M)&=&\int\left(\prod_{j=1}^{L-1}d[W_j] P_j(W_jW_j^\dagger)\right) d\mu(U_L) Q_L(U_L) {\det}^{-n/\widetilde{\gamma}}U_L\nonumber\\
 &&\times {\det}^{1/(\gamma\widetilde{\gamma})}\left[W^{(L-1)}\left(W^{(L-1)}\right)^\dagger\otimes U_L-M\right]\nonumber\\
 &&\hspace*{-1.1cm}=\int\left(\prod_{j=1}^Ld[W_j] P_j(W_jW_j^\dagger)\right)d\mu(U_L) Q_L(U_L) {\det}^{k/\gamma}W^{(L-2)}\left(W^{(L-2)}\right)^\dagger\nonumber\\
 &&\times {\det}^{\frac{1}{\gamma\widetilde{\gamma}}}\left[W_{L-1} W_{L-1}^\dagger\otimes\Id_{\widetilde{\gamma}k}-X_{L-2}^{-1}MY_{L-2}^{-1}\right]\label{part-prod-b}
\end{eqnarray}
where $X_{L-2}=W^{(L-2)}\otimes\sqrt{U_L}$, $Y_{L-2}=(W^{(L-2)})^\dagger\otimes\sqrt{U_L}$, and $Q_L$ is computed as in the projection formula~\eqref{Q-def-b}. This procedure yields a recursion resulting in the following expression for the partition function,
\begin{eqnarray}
 Z_{\Pi}(M)&=&\int \left(\prod_{j=1}^{L}d\mu(U_j) Q_j(U_j) \right){\det}^{-n/\widetilde{\gamma}}U_L\cdots U_1\nonumber\\
 &&\times {\det}^{1/(\gamma\widetilde{\gamma})}\left[\Id_{\gamma n}\otimes\sqrt{U_L}\cdots\sqrt{U_2}U_1\sqrt{U_2}\cdots\sqrt{U_L}-M\right],\label{part-prod-c}
\end{eqnarray}
where each matrix $U_j$ is an element in the circular ensemble $\CtBE(\widetilde{\gamma}k)$.

In the final step we replace $U'_j=\sqrt{U_L}\cdots\sqrt{U_{j+1}}U_j\sqrt{U_{j+1}}\cdots\sqrt{U_L}$ which preserves the symmetries such that $U'_j\in\CtBE(\widetilde{\gamma}k)$. For this purpose we use two facts. First, the Haar measure is invariant under $d\mu(U)=d\mu(VUV^T)$ for all $V\in\U(\widetilde{\gamma}k)$ resulting from the fact that the explicit form of the Haar measure of $\CtBE(\widetilde{\gamma}k)$ is $d\mu(U)\propto {\det}^{-k/\gamma-(\gamma-\widetilde{\gamma})/2}Ud[U]$ with $d[U]$ the product of the differentials of all independent matrix entries \cite{Dys-Circ,superbosonization}. Second, the weights $Q_j$ are also invariant under $Q_j(U)=Q_j(VUV^\dagger)$ for all $V\in\U^{(\beta)}(k)$. Hence these weights have an expression in terms of functions of matrix invariants. With the help of a slight abuse of notation one can say that the weights $Q_j$ satisfy a cyclic permutation symmetry, $Q_j(AB)=Q_j(BA)$ for any two matrices $A,B\in\U(\widetilde{\gamma}k)$.

Finally, we find the result
\begin{equation}
 Z_{\Pi}(M)=\int\left(\prod_{j=1}^{L} d\mu(U'_j) Q_j(U'_jU_{j+1}^{\prime\,-1})\right) {\det}^{-n/\widetilde{\gamma}}U'_1{\det}^{1/(\gamma\widetilde{\gamma})}\left[\Id_{\gamma n}\otimes U'_1-M\right]\label{part-prod-d}
\end{equation}
with $U'_{L+1}=\Id_{\widetilde{\gamma}k}$. This result is surprisingly compact. It also reflects the nature of the original product of matrices which is equivalent to a Mellin-like convolution in a matrix space. Also the dual space exhibits this structure of a Mellin-like convolution.

As an example we calculate the orthogonal polynomials ($k=1$) of a product of $L_{\rm WL}$ complex $\GBE=\chi{\rm GUE}$, Eq.~\eqref{W-L-def}, $L_{\rm CL}$ complex $\LBE={\rm LUE}$, Eq.~\eqref{W-L-def}, and  $L_{\rm J}$ complex $\JBE={\rm JUE}$,~\eqref{W-L-def}. We assume this product to be ordered, i.e. first the Wishart-Laguerre, then the Cauchy-Lorentz, and finally the Jacobi matrices. The result does not depend on this ordering, see the discussion in \cite{IK14}. Then the orthogonal polynomials are
\begin{eqnarray}\label{pol-prod}
 Z_{L_{\rm WL}L_{\rm CL}L_{\rm J}}^{(\beta=2,k=1)}(m\Id_n)&\propto&\oint (1-mz_1^{-1})^n\left(\prod\limits_{j=1}^{L_{\rm WL}}\frac{dz_j}{z_j} \left[\frac{z_{j+1}}{z_j}\right]^{\nu_j}e^{\frac{1}{\Gamma_j^2}\frac{z_j}{z_{j+1}}}\right)\\
 &&\times\left(\prod\limits_{j=L_{\rm WL}+1}^{L_{\rm WL}+L_{\rm CL}}\frac{dz_j}{z_j} \left[\frac{z_{j+1}}{z_j}\right]^{\nu_j}\left[\Gamma_j^2+\frac{z_j}{z_{j+1}}\right]^{\mu'_j-1}\right)\nonumber\\
 &&\times\left(\prod\limits_{j=L_{\rm WL}+L_{\rm CL}+1}^{L_{\rm WL}+L_{\rm CL}+L_{\rm J}}\frac{dz_j}{z_j} \left[\frac{z_{j+1}}{z_j}\right]^{\nu_j}\left[\Gamma_j^2-\frac{z_j}{z_{j+1}}\right]^{-\kappa'_j-1}\right)\nonumber\\
 &&\hspace*{-1.5cm}\propto\sum_{j=0}^{n}\frac{\prod_a\Gamma[n+\kappa_a+\nu+j+1]}{j!(n-j)!(\prod_a(\nu_a+j)!)(\prod_a\Gamma[\mu_a-n-\nu-j])}\left(-\frac{m}{\Gamma^2}\right)^j.\nonumber
\end{eqnarray}
with $z_{L_{\rm WL}+L_{\rm CL}+L_{\rm J}+1}=1$, $\mu'_j=\mu_j-n$, $\kappa'_j=\kappa_j+n$, and $\Gamma^2=\prod_j\Gamma_j^2$. The product of the Gamma functions runs over the possible values for $\nu_a$, $\kappa_a$, and $\mu_a$. The  polynomial~\eqref{pol-prod} is a hypergeometric function and, thus, a Meijer G-function~\cite{ASbook}. It agrees for certain values of the parameters $L_{\rm WL}$, $L_{\rm CL}$, and $L_{\rm J}$ with known results~\cite{Meijer-G-sing-bet-2,KKS15}. What is completely new are the results for $\beta=1,4$ and $k=1$ which are essentially the same polynomials. Here the other approaches failed because of unknown group integrals.

\section{Hard Edge Scaling limit of Product Matrices}\label{Sec:hard}

Up to now every calculation was done for finite $p$ such that we made no approximation and the projection formula was exact. However to make contact to physical systems and universality we have to zoom onto the local scale somewhere of the spectrum. A very prominent scaling is the one to a vicinity around the origin also known as the hard edge scaling limit.

As a simple but non-trivial example, we choose the matrix product of the previous section with the source $M=\widetilde{\gamma}(\prod_j\Gamma_j^2)\Id_{\gamma n}\otimes\widehat{m}/[n(\prod_a(\mu_a-n/\widetilde{\gamma}))(\prod_a(\kappa_a+n/\widetilde{\gamma}))]$. In particular we consider the scaling limit $n\to\infty$ and $\nu_j$, $\widehat{\mu}_j=(\mu_j/n-1/\widetilde{\gamma})$, $\widehat{\kappa}_j=(\kappa_j/n+1/\widetilde{\gamma})$, and $\widehat{m}$ fixed. Then one can easily show that the asymptotics of each weight, regardless what kind of random matrix we consider, is
\begin{equation}\label{asy-Q}
 Q_j(\alpha U)\overset{n\gg1}{\propto}{\det}^{\nu_j/\widetilde{\gamma}}Ue^{\tr U}
\end{equation}
with $\alpha=\Gamma_j^2$ for $\GBE$, $\alpha=\Gamma_j^2/(n\widehat{\mu}_j)$ for $\LBE$, and $\alpha=\Gamma_j^2/(n\widehat{\kappa}_j)$ for $\LBE$. After a proper rescaling of the matrices $U_j$ the partition function~\eqref{part-prod-d} takes the asymptotic form
\begin{equation}
 Z_{\Pi}(M)\overset{n\gg1}{\propto}\int\left(\prod_{j=1}^{L} d\mu(U'_j){\det}^{\nu_j/\widetilde{\gamma}}U_j\right) e^{\tr U_L+\sum_{j=1}^{L-1}\tr U_jU_{j+1}^{-1}-\tr\widehat{m}U_1^{-1}}\label{part-prod-asy}
\end{equation}
 with $L=L_{\rm WL}+L_{\rm CL}+L_{\rm J}$. We underline that no saddlepoint approximation is needed for this limit. Hence the matrices $U_j$ are still elements of the circular ensemble $\CtBE(\widetilde{\gamma}k)$.
 
 For $\beta=2$ the partition function~\eqref{part-prod-asy} yields the Meijer G-kernel of a product of matrices drawn from $\GBE$s, cf. \cite{Meijer-G-sing-bet-2}. This can be seen by diagonalizing the unitary matrices, applying the Itzykson-Zuber integral~\cite{IZ80} and finally integrating over a determinantal point process. The entries of the resulting determinant are Meijer G-functions. Our result emphasizes the conjecture that also this kernel is universal. Indeed we could also have chosen another scaling which still leads to a hard edge scaling limit. Then we would get finite rank deformations of the result~\eqref{part-prod-asy} which was recently discovered for a product of truncated unitary matrices in \cite{KKS15}. Nevertheless the limiting kernel is still a Meijer G-kernel but with other parameters.
 
 From a physical point of view can one ask for the non-linear $\sigma$-model corresponding to the partition function~\eqref{part-prod-asy}. In this framework the function in the exponential function is identified as the potential. The integration domain $\CtBE^L(\widetilde{\gamma}k)$ is the coset of the ``flavour" group which keeps the ``massless Lagrangian'' ($\widehat{m}=0$) in the full theory at finite ``volume'' $n$ invariant divided by the group which keeps the ground state invariant. As in the case $L=1$ the theory is spontaneously broken. For a product matrix the  ``flavour" symmetry at finite ``volume'' $n$ is $\U^L(\widetilde{\gamma}k)$ for $\beta=1,4$ and $[\U(k)\times\U(k)]^L$ for $\beta=2$ which can be readily checked by linearising the product $W^{(L)}$ in the matrices $W_j$. This group is spontaneously broken to $[\U^{(\widetilde{\beta})}(k)]^L$ and the source term for its condensate is the ``mass'' $\widehat{m}$. This non-linear $\sigma$ model generalizes the one for the Wishart-Laguerre ensemble which were found in QCD \cite{QCD-Lag} and mesoscopic systems~\cite{Zirn}. Especially the coupling between different $U_j$ is reminiscent but not completely the same as the recently proposed chiral Lagrangian for high density QCD~\cite{KW14}.

\section{Conclusions}\label{Sec:conclusio}

We briefly presented the projection formula~\cite{KKG14} for averages over products of characteristic polynomials which is a short cut of the supersymmetry method~\cite{SUSY-method,superbosonization}. The general results found by this approach were demonstrated in the case of Wishart-Laguerre ($\GBE$), Cauchy-Lorentz ($\LBE$), and Jacobi ($\JBE$) ensembles, in particular we rederived the corresponding orthogonal polynomials for $\beta=2$. These polynomials are essentially the same when averaging over one characteristic polynomial for $\beta=1$ and over a square root of a characteristic polynomial for $\beta=4$.

Moreover we generalized the projection formula to products of matrices. Since the projection formula works in a unifying way for all three Dyson indices $\beta=1,2,4$ this approach is an ideal alternative compared to other methods like orthogonal polynomials and free probability when studying real or quaternion matrices.  Note that up to now free probability only applies to global spectral properties and to use orthogonal polynomials we need to know group integrals like the Itzykson-Zuber integral~\cite{IZ80} or its polynomial counterpart~\cite{group-int,KKS15}. The projection formula circumvents this problem. In particular we were able to show that the spectral statistics at the hard edge are the same for products of completely different random matrices only depending on the number of matrices defined and their indices $\nu_1,\ldots,\nu_l$ encoding the rectangularity of the matrices. This was done for all three cases $\beta=1,2,4$ and underlines the strength of the projection formula where other methods fail. In the complex case ($\beta=2$) we easily deduce from our results those for the Meijer G-ensembles studied in \cite{Meijer-G-sing-bet-2,KKS15}.

The projection formula also enabled us to identify the non-linear $\sigma$-models and the symmetry breaking pattern for product matrices and derived the potential of the Goldstone manifold. This result is completely new and shows what the effective theory associated to such a product matrix would look like. In particular one can understand a  product matrix by itself as a discrete one-dimensional system. Therefore our results show one way to generalize the zero-dimensional RMT to a one-dimensional theory. Indeed via the DMPK equation such a link to a one-dimensional system was established \cite{been}, though there a different limit is considered.

\section*{Acknowledgements}

I acknowledge partial financial support by the Alexander von Humboldt foundation. Furthermore, I thank G. Akemann, Z. Burda, T. Guhr, J. R. Ipsen, V. Kaymak, M. A. Nowak, and J. J. M. Verbaarschot for fruitful discussions.

\end{document}